\documentclass[aps,pra,twocolumn]{revtex4-1}

\usepackage{amssymb,amsmath}
\usepackage{stix}
\usepackage[colorlinks,linkcolor=blue,citecolor=blue,urlcolor=blue]{hyperref}
\usepackage{graphicx}
\usepackage[utf8]{inputenc}
\usepackage[T1]{fontenc}

\usepackage[dvipsnames]{xcolor}

\renewcommand{\vec}[1]{\mathbf{#1}}
\newcommand{\gvec}[1]{\boldsymbol{#1}}
\newcommand{\dd}{\mathrm{d}}
\newcommand{\en}{\varepsilon}
\newcommand{\iu}{\mathrm{i}}

\graphicspath{{./Figs/}}
\mathchardef\mhyphen="2D \newcommand{\nn}{\nonumber}

\usepackage{chemformula}
\usepackage{dcolumn}
\begin{document}

\title{Spectral properties from Matsubara Green's function approach\,---\,application to
  molecules}

\author{M. Sch{\"u}ler}\email[]{michael.schueler@unifr.ch}
\affiliation{Department of Physics, University of Fribourg, 1700
  Fribourg, Switzerland}
\author{Y. Pavlyukh} 
\affiliation{Department of Physics and Research Center OPTIMAS, University of Kaiserslautern,
  P.O. Box 3049, 67653 Kaiserslautern, Germany
}
\affiliation{Institut f\"{u}r Physik, Martin-Luther-Universit\"{a}t Halle-Wittenberg,
  06099 Halle, Germany}
\begin{abstract}
  We present results for many-body perturbation theory for the one-body Green's function
  at finite temperatures using the Matsubara formalism. Our method relies on the accurate
  representation of the single-particle states in standard Gaussian basis sets, allowing
  to efficiently compute, among other observables, quasiparticle energies and Dyson
  orbitals of atoms and molecules. In particular, we challenge the second-order treatment
  of the Coulomb interaction by benchmarking its accuracy for a well-established test set of
  small molecules, which includes also systems where the usual Hartree-Fock treatment
  encounters difficulties. We discuss different schemes how to extract quasiparticle
  properties and assess their range of applicability. With an accurate solution and
  compact representation, our method is an ideal starting point to study electron dynamics
  in time-resolved experiments by the propagation of the Kadanoff-Baym equations.
\end{abstract}

\maketitle

\section{Introduction}

Many-body perturbation theory (MBPT) is one of the most important tools for the prediction
of electronic structures from first principles~\cite{mahan_many-particle_2000}.  The
controllability of approximations derived from diagrammatic techniques, the wealth of
information about the spectroscopic observables contained in the single-particle Green's
function, and the compatibility of the method with the time-propagation and the
description of transport properties are believed to be the strong points of the Green's
function approach. However, technical realization of these advantages has proved
difficult.

\begin{figure}[t]
  \centering
  \includegraphics[width=0.6\columnwidth]{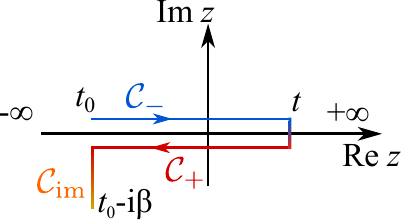}
  \caption{(Color online) Time-arguments of the electron Green's function belong to
      the general contour $\mathcal{C}$ consisting of the forward branch $\mathcal{C}_-$
      on the real axis, the backward branch $\mathcal{C}_+$ and the imaginary branch
      $\mathcal{C}_\mathrm{im}$. The arrows indicate the direction of the
      contour-ordering. $\beta$ denotes the inverse temperature. \label{fig:contour}}
\end{figure}

In this work we focus on the extraction of spectral information encoded in the Matsubara
Green's function and on the benchmarking of a popular second Born approximation (2BA) for
the self-energy. This study is motivated by the fact that the solution of the Dyson
equation on the imaginary time-track is the first step of a typical nonequilibrium Green's
function (NEGF) approach in the two-times
plane~\cite{balzer_nonequilibrium_2012,stefanucci_nonequilibrium_2013}. The power of the
NEGF approach has been demonstrated by the description of ultra-fast carrier dynamics
\cite{haug_interband_1992,vu_time-dependent_2000,vu_relaxation_2006,eckstein_photoinduced_2013},
time-resolved photoemission
\cite{kemper_mapping_2013,sentef_examining_2013,kemper_effect_2014,schuler_time-dependent_2016}
and photoionization of atoms and molecules
\cite{balzer_time-dependent_2010,balzer_efficient_2010}. Some of these results were
obtained by starting from the non-interacting reference state and switching on the
interaction adiabatically. The numerical scheme simplifies significantly in this
case. However, it requires the propagation up to longer times, increasing the computer
memory requirements and the computational time considerably~\cite{balzer_electronic_2012}.
Therefore, an efficient NEGF solver necessarily incorporates the vertical track of the
Keldysh contour~(Fig.~\ref{fig:contour}) in the propagation scheme~\footnote{The
  generalized Kadanoff-Baym ansatz~\cite{lipavsky_generalized_1986,latini_charge_2014} is
  another way to alleviate this difficulty}.

Our choice of an approximation for the self-energy\,---\,the second Born
approximation\,---\,has been shown to be relevant for molecular
system~\cite{balzer_time-dependent_2010,balzer_efficient_2010,perfetto_first-principles_2015}.
In particular, exchange effects (the 2BA includes exchange up to second order) are known
to have a significant contribution to the total energy
~\cite{marsman_second-order_2009,schafer_quartic_2017}\,---\,a fact that is elevated for
molecular systems.  In contrast to the broadly used (for equilibrium calculations) $GW$
~\cite{pavlyukh_life-time_2004,stan_levels_2009,van_setten_gw-method_2013,
  van_setten_gw100:_2015,kaplan_quasi-particle_2016} and the $T$-matrix approximations
\cite{von_friesen_successes_2009,puig_von_friesen_kadanoff-baym_2010,
  romaniello_beyond_2012}, it possesses an additional benefit of the time-locality. By
this we mean that the self-energy $\Sigma(z_1,z_2)$ is a functional of the Green's
function with the same time-arguments. This simplifies the time-propagation considerably.

Many-body approximations have been tested extensively in the energy domain and used as the
initial step for the
time-propagation~\cite{von_friesen_successes_2009,balzer_efficient_2010,
  balzer_time-dependent_2010,hopjan_merging_2016}.  Direct construction of the initial
propagator on the Keldysh contour is less common.  Extensive study of the \emph{spectral
  properties} of the 2BA theory have been performed on finite lattice systems based on the
Pariser–Parr–Pople model~\cite{sakkinen_kadanoffbaym_2012}. 2BA shows a clear improvement
over the Hartree-Fock (HF) electronic structure: predicting in accordance with the exact
diagonalization results the correlation-induced satellites, and yielding correct shifts of
spectral features as a function of the interaction strength. However, oscillatory
noise-like features, as well as broadening of the peaks in the frequency domain appear due
to the finite propagation length in time-domain. How spectral features are reproduced in
realistic systems, and how 2BA compares with standard quantum chemistry methods are still
two open prominent questions. Some steps in answering these questions have been undertaken
recently focusing on the \emph{integral properties}. The importance of proper frequency
grid for the representation of Matsubara Green's function (MGF) has been demonstrated by
Kananenka \emph{et al.} in a study of simple molecular
systems~\cite{kananenka_efficient_2016-1}. They proposed a method based on spline
interpolation and established a criterion determining the accuracy of the
results. However, here we focus on the intrinsic limitations of the 2BA rather than on the
errors induced by the numerical procedure.

In addition to already mentioned restrictions induced by the finite grid representation of
the MGF, the extraction of the spectral properties from the imaginary time propagation is
a nontrivial mathematical problem. The maximum entropy and the generalized Pad\'{e}
approximation have been compared with the direct Laplace transform by Dirks \emph{et
  al.}~\cite{dirks_extracting_2013}. While on the conceptual level the former method is
superior, it was suggested that the actual performance for realistic systems is rather
subtle depending on the choice of the self-energy approximation. This is our second
motivation for the benchmarking of various approaches for small molecules from the well
established test sets and comparing the performances of the 2BA and the coupled-cluster
approach. In this work, we analyze the closed-shell neutral molecules of the G2-1 set and
the non-hydrogenic molecules from the G2/97
set~\cite{curtiss_assessment_1997,curtiss_assessment_1998,haunschild_new_2012}.  The
latter offers the advantage of directly comparing the 2BA to the $GW$ approximation from
ref.~\onlinecite{pham_$gw$_2013}. The G2-1 test set also contains a number of molecules
(such as the dimers Li$_2$, F$_2$, Na$_2$ and P$_2$) for which the electronic structure
within the Hatree-Fock treatment differs considerably from the accurate coupled-cluster
results. Thus, the limits of the 2BA as such and the extraction of quasiparticle
properties can be assessed in an unbiased way.

As the last ingredient of this study we consider the extraction of the spectral
information from the MGF using the extended Koopmans' theorem
(EKT)~\cite{chipman_methods_1977}. While extensive tests of this possibility have been
performed in the quantum chemistry framework~\cite{vanfleteren_exact_2009}, there has also
been proposals to use EKT within the Green's function
approach~\cite{dahlen_self-consistent_2005,stan_fully_2006}. However, the EKT cannot be
considered an equivalent substitute of the aforementioned spectral methods. While they use
in principle all dynamical information encapsulated in the MGF, the latter approach solely
relies on the one and two-particle density matrices and is sensitive to the asymptotic
behavior of the bound state wave-functions~\cite{katriel_asymptotic_1980}. Because of this
restriction, only the first ionization potential and electron affinity in each symmetry
class can be obtained. However, the advantage is that EKT can be applied to any correlated
ground state, e.~g., from coupled cluster approaches.

The work is organized as follows. n Sec.~\ref{sec:methods} we
summarize the well known self-energy expressions specializing on the
Matsubara formalism, recall basic facts about the EKT, the analytic
continuation (AC) and the Pad\'{e} approximation, and describe the
numerical implementation of these methods for molecular systems. In
Sec.~\ref{sec:g2} the results of benchmark calculations are presented
and compared with reference experimental and coupled cluster numerical
data. In Sec.~\ref{sec:summary} we discuss in details our main finding
that the 2BA can compete with accurate quantum chemistry methods and
thus endorse the method as accurate and extendible approach to
equilibrium and excited-state properties of molecules.
\section{Methods\label{sec:methods}}
Our calculations are performed in the molecular orbital basis. Its size will be denoted as
$N_\text{bas}$. Correspondingly, the MGF
$\vec{G}(\tau)$ and self-energy $\vec{\Sigma}(\tau)$ are matrices related by the Dyson
equation
\begin{equation}
  \label{eq:MDE1}
  \vec{G}(\tau) = \vec{g}(\tau) + \int^0_{-\beta}\!\dd \tau_1\!\int^0_{-\beta}\!\dd \tau_2
  \,\vec{g}(\tau-\tau_1)\vec{\Sigma}(\tau_1-\tau_2)\vec{G}(\tau_2) \ .
\end{equation}
Here $\vec{g}(\tau)$ is the reference Green's function 
\begin{equation}
  g_{ij}(\tau) = \delta_{ij} \left[ (n_i-1)\theta(\tau)+ n_i \theta(-\tau) \right]
  e^{-(\en_i-\mu)\tau},
\end{equation}
with $n_i = n_\mathrm{F}(\en_i-\mu)$ ($n_\mathrm{F}(\omega)\equiv(1+e^{\beta
  \omega})^{-1}$ denotes the Fermi distribution function), and $\en_i$ standing for the HF
eigenvalues. Introducing the Coulomb matrix elements
\begin{align}
\left( i l | k j \right)&= \int\!\dd \vec{r}\!\!\int\!\dd \vec{r}^\prime \,
\phi^*_i(\vec r) \phi^*_j(\vec r^\prime) \mathscr{v}(\vec r - \vec{r}^\prime)
\phi_k(\vec r^\prime) \phi_l(\vec r),
 \nonumber \ 
\end{align}
the constituent second order self-energy can be efficiently computed using the matrix
multiplication:
\begin{align}
  \label{eq:sigmar2bb1}
  \Sigma^{(2)}_{ij}(\tau) = \sum_{klmnpq} &\left( ik| mq \right) \left( l j | p n\right)
   \Big[2 G_{kl}(\tau) G_{mn}(\tau)G_{pq}(-\tau) \nn \\
    &\quad - G_{kq}(\tau)G_{mn}(\tau)G_{pl}(-\tau) \Big] \ ,
\end{align}
While this does not change the complexity proportional to the fifth power of the
number of basis functions ($\mathcal{O}(N_\text{bas}^5)$), the use of specialized
libraries and parallelization allows to achieve a substantial speed up (the brute force
approach leads to $\mathcal{O}(N_\text{bas}^8)$
scaling~\cite{balzer_electronic_2012}). Another possibility to increase the performance
would be to use the finite-element discrete variable representation, which was shown to
lead to $\mathcal{O}(N_\text{bas}^4)$ scaling~\cite{balzer_efficient_2010}.

The Dyson equation~\eqref{eq:MDE1} is typically~\cite{fetter_quantum_2003} solved by
Fourier-transforming $\vec{G}(\tau)$ to imaginary frequencies $\omega_m$,
\begin{align}
  \label{eq:fourier1}
  \widetilde{\vec{G}}(\iu \omega_m) = \int^\beta_0\!\dd \tau \,\vec G(\tau)e^{\iu
    \omega_m\tau}, \quad\omega_m=\frac{(2m+1)\pi}{\beta} \ ,
\end{align}
yielding an algebraic Dyson equation. The self-energy is, however, most efficiently
evaluated in $\tau$-space. Hence, switching back and forth between time and frequency
representation is the standard implementation of the self-consistency
cycle. Due to the non-continuous behavior of the MGF at $\tau=0$, the Fourier coefficients
$\widetilde{\vec{G}}(\iu \omega_m)$ behave as $(\iu \omega_m)^{-1}$ for large $|m|$. This
slow convergence introduces significant numerical errors which are countered by tail
corrections. However, the standard first-order correction
scheme~\cite{aoki_nonequilibrium_2014} still requires a typical number of thousands of
frequencies $\iu \omega_m$ to achieve accurately converged results. Higher-order tail
corrections~\cite{kananenka_efficient_2016-1} is a promising perspective to improve the
efficiency of this  scheme.

An alternative approach is to solve the Dyson equation directly as integral equation.  By
replacing the integration over the imaginary time arguments by a suitable quadrature with
points $\tau_p$ and weights $w_p$, the integral equation~\eqref{eq:MDE1} is recasted into
a system of linear equations~\cite{balzer_nonequilibrium_2009}:
\begin{align}
  \label{eq:MDE2}
  \sum^{N_\mathrm{quad}}_{q=1} \left[\vec{I} \delta_{pq} - w_q
    \vec{Z}(\tau_p,\tau_q) \right]\vec{G}(\tau_p) = \vec{g}(\tau_p)
\end{align}
with integral kernel
\begin{align}
  \label{eq:Zkern}
  \vec{Z}(\tau,\tau^\prime) = \int^0_{-\beta}\!\dd \tau^{\dprime}\,
  \vec{g}(\tau-\tau^{\dprime})\vec{\Sigma}(\tau^{\dprime}-\tau^\prime) \ .
\end{align}
Combining $N_\mathrm{bas}$ basis and $N_\mathrm{quad}$ grid indices into a multi-index,
Eq.~\eqref{eq:MDE2} is transformed into a $N_\mathrm{bas} N_\mathrm{quad}$-dimensional
system of linear equations with $N_\mathrm{bas}$ right-hand sides.  Solving the Dyson
equation directly as integral equation yields a solution free of high-frequency
artefacts. By evaluating eq.~\eqref{eq:MDE2} and \eqref{eq:Zkern} by higher-order
quadrature schemes~ \footnote{We employ adaptive Gaussian quadrature for the
  kernel~\eqref{eq:Zkern} and fourth-order weights for eq.~\eqref{eq:MDE2} (including
  boundary corrections). The global error of our solution scheme scales as
  $\mathcal{O}(\Delta \tau^4)$, where $\Delta \tau$ is the typical grid spacing.}, we
obtain a highly accurate solution even for moderate number of grid points $\tau_p$. The
numerical bottleneck of the method is the additional computational cost of constructing
the kernel~\eqref{eq:Zkern}. However, its calculation can be efficiently incorporated into
a distributed memory scheme for solving the linear equation~\eqref{eq:MDE2}, giving rise
to excellent scaling with the number of processing cores. We remark that the MGF obtained
by the solving in the frequency space can provide a good initial guess for
$\vec{G}(\tau)$ to be inserted in the right-hand side of eq.~\eqref{eq:MDE1}. Constructing
an improved MGF from the left-hand side and substituting back into the convolution on the
right-hand side constitutes an iterative solution of the Dyson
equation~\cite{schuler_nonthermal_2017}.

\begin{figure}[t]
  \centering
  \includegraphics[width=\columnwidth]{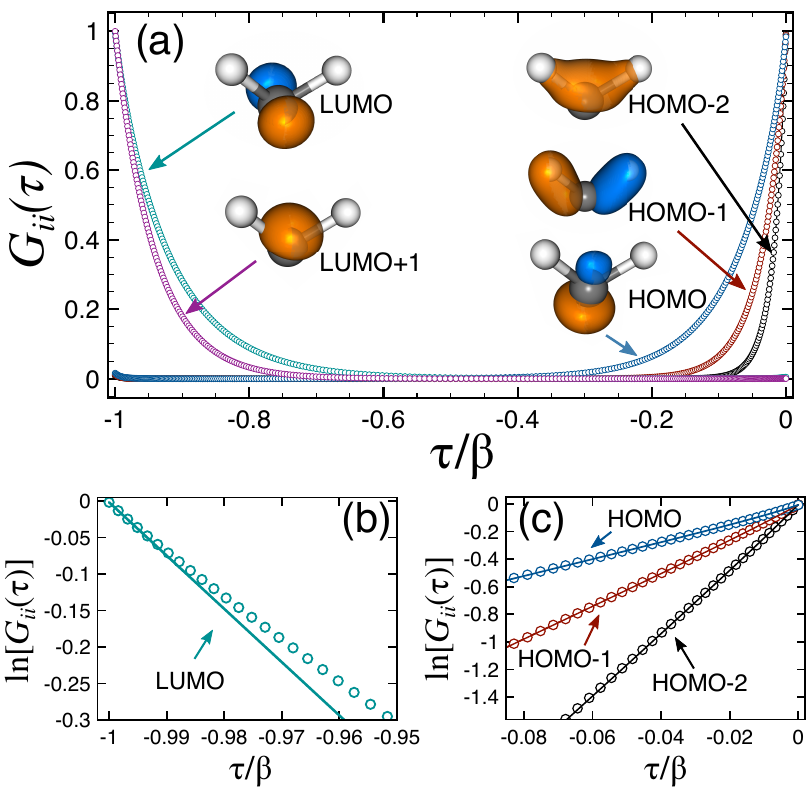}
  \caption{(Color online) (a) Typical behavior of MGF as a function of imaginary time
    $-\beta<\tau<0$, $G_{ik}(\tau)=-G_{ik}(\tau+\beta)$ for the CH$_2$ molecule as an
    example. The circles represent the grid points used in the calculations. The insets
    depict the molecular orbitals corresponding to the diagonal matrix elements
    $G_{ii}(\tau)$. On the panels below, we compare the MGF obtained by solving the
      Dyson equation using the HF self-consistent procedure (circles) and the MGF
      reconstructed from energies and states of the EKT eigenproblem~\eqref{eq:EKT} (full
      lines) for the LUMO (b) and the occupied valence orbitals (c).
    \label{fig:HOMO-LUMO}}
\end{figure}

In order to deal with cusps at the boundaries $\tau=-\beta$ and $\tau=0$ while retaining a
compact representation of the MGF, we employ a grid $\tau_p$ with exponentially increased
density at the boundaries (Fig.~\ref{fig:HOMO-LUMO}(a)). The exponential scaling is
optimized to best represent the (noninteracting) MGF corresponding to highest occupied
molecular orbital (HOMO). For achieving converged results, we typically need
$N_{\mathrm{quad}}=200$ to $N_{\mathrm{quad}}=300$ grid points.

We have implemented three different levels of self-consistency at which the Matsubara GF
is determined:
\begin{itemize}
\item[(i)] the non-self-consistent (non-sc) treatment, where the Dyson equation is solved
  only once using the self-energy constructed from the reference Hartree-Fock (HF) Green's function;
\item[(ii)] the partially self-consistent scheme where only the mean-field part of the
  Hamiltonian is updated (HF-sc) until the convergence of the MGF;
\item[(iii)] the fully self-consistent scheme (full-sc), where the Dyson equation
  is solved and the self-energy is constructed repeatedly until the convergence of the
  MGF is achieved.
\end{itemize}
The convergence is achieved when the norm of deviation of
$\vec G(\tau)$ between subsequent iteration steps for all imaginary
time arguments $\tau_p$ is below a specified threshold. This criterion
is more stringent than the convergence of the density matrix, which
corresponds to the value of MGF at $\tau=0$.

Direct extraction of the spectral function $A_{ik}(\omega)$ from the MGF amounts to
solving the integral equation
\begin{align}
\label{eq:gtauspectral}
\vec G(\tau) &=\int \!\frac{\dd\,\omega}{2\pi} \vec{A}(\omega)\frac{e^{-\tau\omega}}{1+e^{-\beta
    \omega}},
\end{align}
which is a nontrivial task~\cite{dirks_extracting_2013}. That is why we explore the
AC and the EKT routes.
\paragraph{Analytic continuation} transforms the Green's function of the imaginary time
argument $\tau$ into the function of complex frequency $\zeta$ in a sequence of two steps
$\vec G(\tau)\rightarrow\widetilde{\vec{G}}(\iu \omega_m)\rightarrow
\widetilde{\vec{G}}(\zeta)$. For $\zeta$ in the vicinity of real axis, the latter quantity
relates to retarded/advanced GFs ($\widetilde{\vec{G}}(\omega \pm \iu \eta) =
\vec{G}^\mathrm{R/A}(\omega + \mu)$) and yields the spectral function according to
\begin{align*}
  \vec A(\omega) &= 
  \iu \left[\widetilde{\vec{G}}(\omega-\mu+\iu \eta) -\widetilde{\vec{G}}(\omega-\mu-\iu
    \eta) \right]_{\eta\rightarrow 0^+} \ .
\end{align*}

For equidistant grids in imaginary time, fast Fourier transformation to the imaginary
frequency domain is the standard procedure. However, for our efficient solution scheme of
the Dyson equation\,---\,which relies on an optimized non-equidistant grid of
$\tau$-points\,---\,it is more efficient to employ an orthogonal polynomial
representation~\cite{kananenka_efficient_2016}.  The Fourier coefficients of the Matsubara
GF~\eqref{eq:fourier1} are computed by representing the function in terms of Legendre
polynomials $P_n(x)$:
\begin{align}
  \vec G(\tau) = \frac{1}{\beta}\sum^\infty_{n=0} \sqrt{2n+1}\,\,P_n\left(\frac{2\tau}{\beta}-1\right)\vec C_n \ ,
\end{align}
yielding~\cite{boehnke_orthogonal_2011}
\begin{align}
  \widetilde{\vec{G}}(\iu \omega_m) = (-1)^m \sum^\infty_{n=0}
  \iu^{n+1}\sqrt{2n+1}\,\,j_n\left(\frac12\beta\omega_m\right) \vec C_n\ .
\end{align}
Here $j_n(x)$ denotes the spherical Bessel function of the first kind. On the
second step, the complex function $\widetilde{\vec{G}}(\zeta)$ is represented by its
Pad\'{e} approximant constructed from the points $\iu \omega_m$. In
practice, the order of the Legendre polynomials is truncated at $\simeq 64$, yielding
excellent accuracy. The order of the Pad\'{e} approximation (we
choose 28) plays
only a minor role.

\paragraph{Extended Koopmans' theorem}
One quantity immediately available from MGF is the density matrix $\gvec \gamma^\pm =\vec
\lim_{\tau\rightarrow 0^\pm}\vec G(\tau)$ (upper/lower sign for particle/hole density,
respectively).  Computation of the quasiparticle excitations additionally requires the
two-body correlation function as encoded in the first derivative $\vec \Delta^\pm =
-\lim_{\tau\rightarrow 0^\pm}\partial_\tau \vec
G(\tau)$~\cite{dahlen_self-consistent_2005,stan_fully_2006}. With these two ingredients a
generalized eigenvalue problem
\begin{align}
  \label{eq:EKT}
  \vec \Delta^\pm \vec u^\pm_{\alpha} = \epsilon^\pm_\alpha\gvec{\gamma}^\pm\vec
  u^\pm_{\alpha}
\end{align}
yields the quasiparticle energies $\epsilon^\pm_\alpha$. Corresponding
Dyson orbitals can be obtained from the normalized
($\left[\vec u^\pm_{\alpha}\right]^\dagger \gvec{\gamma}^\pm\vec
u^\pm_{\beta}=\delta_{\alpha,\beta}$) eigenvectors as follows
$\gvec{\phi}^{\pm}_\alpha=\gvec{\gamma}^\pm \left(\vec
  u^\pm_{\alpha}\right)^*$. In terms of $\epsilon^\pm_\alpha$ and
$\gvec{\phi}^{\pm}_\alpha$, the spectral function is given by:
\begin{align}
\label{eq:deltaspectral}
\vec A(\omega)=2\pi\sum_{i=\pm}\sum_\alpha \left[\gvec{\phi}^{i}_\alpha\right]^\dagger
\gvec{\phi}^{i}_\alpha \delta(\omega-\epsilon^i_\alpha).
\end{align}

Let us remark on the relation of the AC and the EKT. Provided one has
found an exact MGF (by exact diagonalization, for instance), the EKT reproduces (in the
limit $\beta \rightarrow \infty $) the exact many-body energies. The same is true for
the AC. At the level of finite-order MBPT, the relation is less
clear. Low-order diagrammatic methods such as the 2BA or the $GW$ approximation result in
additional features like satellites and broadening, which can not be captured by the
EKT. A typical example where the simple exponential behavior of the MGF implied by the EKT
(the Green's function is reconstructed by substituting the spectral
function~\eqref{eq:deltaspectral} into the integral representation
Eq.~\eqref{eq:gtauspectral}) deviates from the self-consistent solution is shown in
Fig.~\ref{fig:HOMO-LUMO}(b)--(c) for the lowest unoccupied molecular orbital (LUMO) and
the occupied orbitals. One can expect that in case the effects of the 2BA is primarily
given by shifting the HF energies, the EKT and resulting peaked spectral
function~\eqref{eq:deltaspectral} is an excellent approximation (as can be seen for the
occupied orbitals), while it might give inconsistent results if the above mentioned
features of MBPT come into play. For this reason, we employ both methods for obtaining
quasiparticle properties and compare them.

\begin{figure}[t]
  \centering
  \includegraphics[width=\columnwidth]{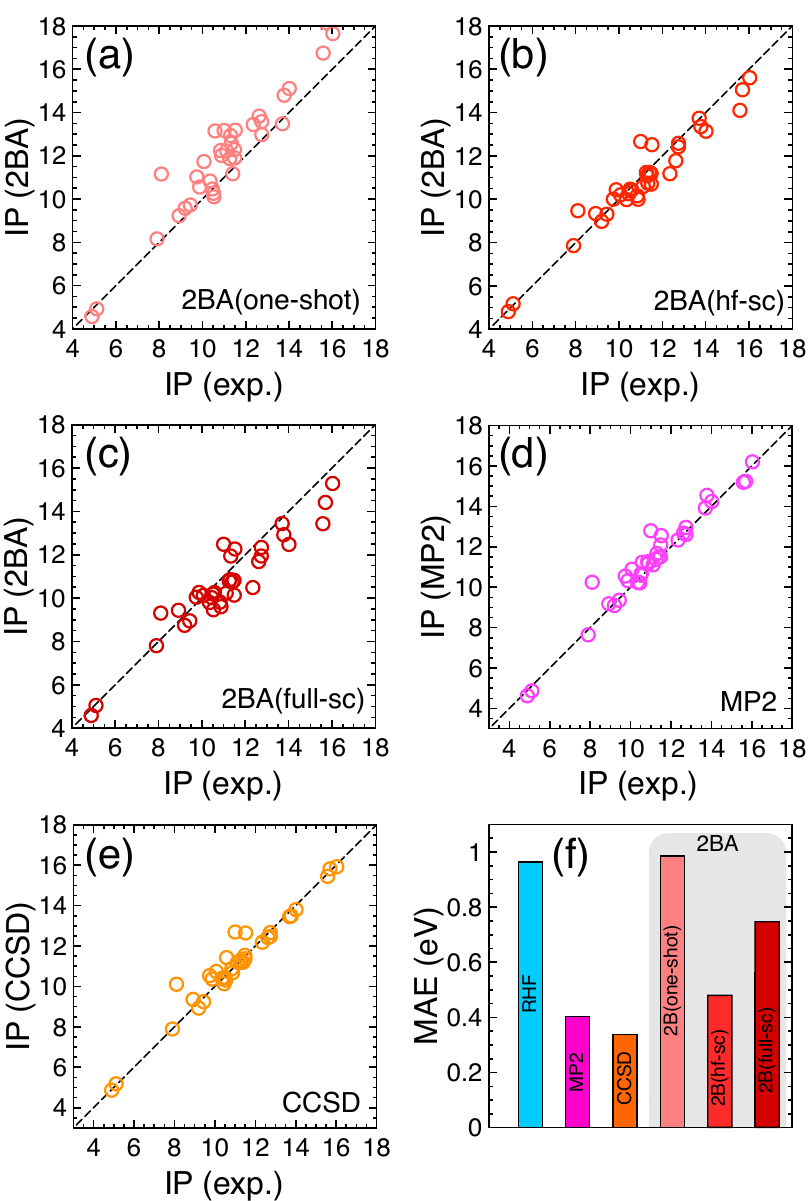}
  \caption{(Color online) Ionization potentials of the G2-1 molecules (a)--(c) within the
    2BA (using the EKT) and, for comparison, (d) within MP2 and (e) CCSD, vs. the
    experimental values (taken from ref.~\cite{johnson_nist_nodate}). Values are in
    eV. Panel (f) shows the mean absolute error (MAE) for each method with respect to the
    experimental values; color coding is consistent with other
    panels.  \label{fig:ip_vs_exp}}
\end{figure}

\section{Electronic properties of G2 molecules\label{sec:g2}}
For a comprehensive benchmark, we study all 36 neutral closed-shell molecules from the
G2-1 test set~\cite{curtiss_assessment_1997} with geometries optimized at the B3LYP/6-31*
level. The restricted HF calculation is performed using the aug-cc-pVDZ
basis~\cite{noauthor_emsl_nodate,feller_role_1996} as the starting point, all the matrix
elements are transformed from the atomic to molecular orbital basis using our in-house
code~\cite{pavlyukh_electron_2013}. In this basis the Dyson equation~\eqref{eq:MDE1} is
subsequently solved for the low-temperature case $\beta=80$ and spectral properties are
determined.  
We tested the convergence of the results with respect to the basis size by introducing a
cutoff energy $E_\mathrm{cut}$ such that $\en_i<E_\mathrm{cut}$ for all states $i$. In
general, higher molecular orbitals are not described well by the Gaussian basis set, such
that including them leads to additional errors. On the other hand, the molecular basis set
needs to be large enough to describe adding an extra electron. We performed calculations
for two values for the cutoff: $E_\mathrm{cut}=0.5$~a.\,u. and
$E_\mathrm{cut}=1.5$~a.\,u., respectively. In what follows, we present the results for
$E_\mathrm{cut}=1.5$~a.\,u., while the corresponding results for the smaller cutoff are
summarized in Appendix~\ref{sec:app:G2-1}.

For comparison, the second-order M\o{}ller-Plesset perturbation theory (MP2) and the
coupled-cluster method including single and double excitations (CCSD) is used. With these
two reference methods, the total energies of the neutral and the positively/negatively
charged ions were computed using the same basis set (aug-cc-pVDZ) and the active
  space, yielding accurate estimates to the vertical ionization potential (IP) and the
electron affinity (EA) according to the energy difference method. It should be noted,
however, that for some molecules the underlying HF calculation suffers from
multi-configuration instabilities. In such cases, the HF ground state of the neutral or
ionized system differs significantly from the true electronic state. We will come back to
this point later.

\subsection{Ionization potentials and electron affinities of G2-1 molecules\label{sec:IP:EA}}

\begin{figure}[t]
  \centering
  \includegraphics[width=\columnwidth]{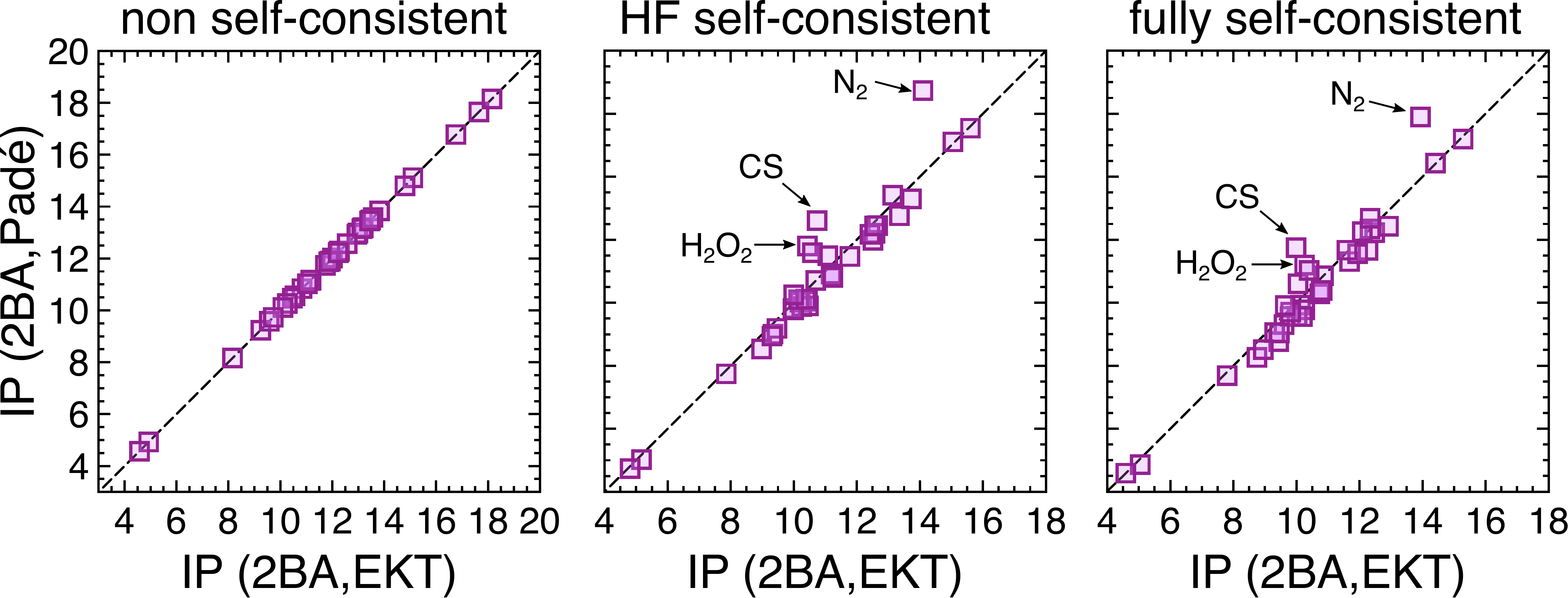}
  \caption{(Color online) Comparison of the ionization potentials obtained from the EKT
    and the Pad\'{e} analytic continuation for three different levels of
    self-consistency. Values are in eV. \label{fig:ip_pade_vs_ekt}}
\end{figure}

\begin{figure}[b]
  \centering
  \includegraphics[width=\columnwidth]{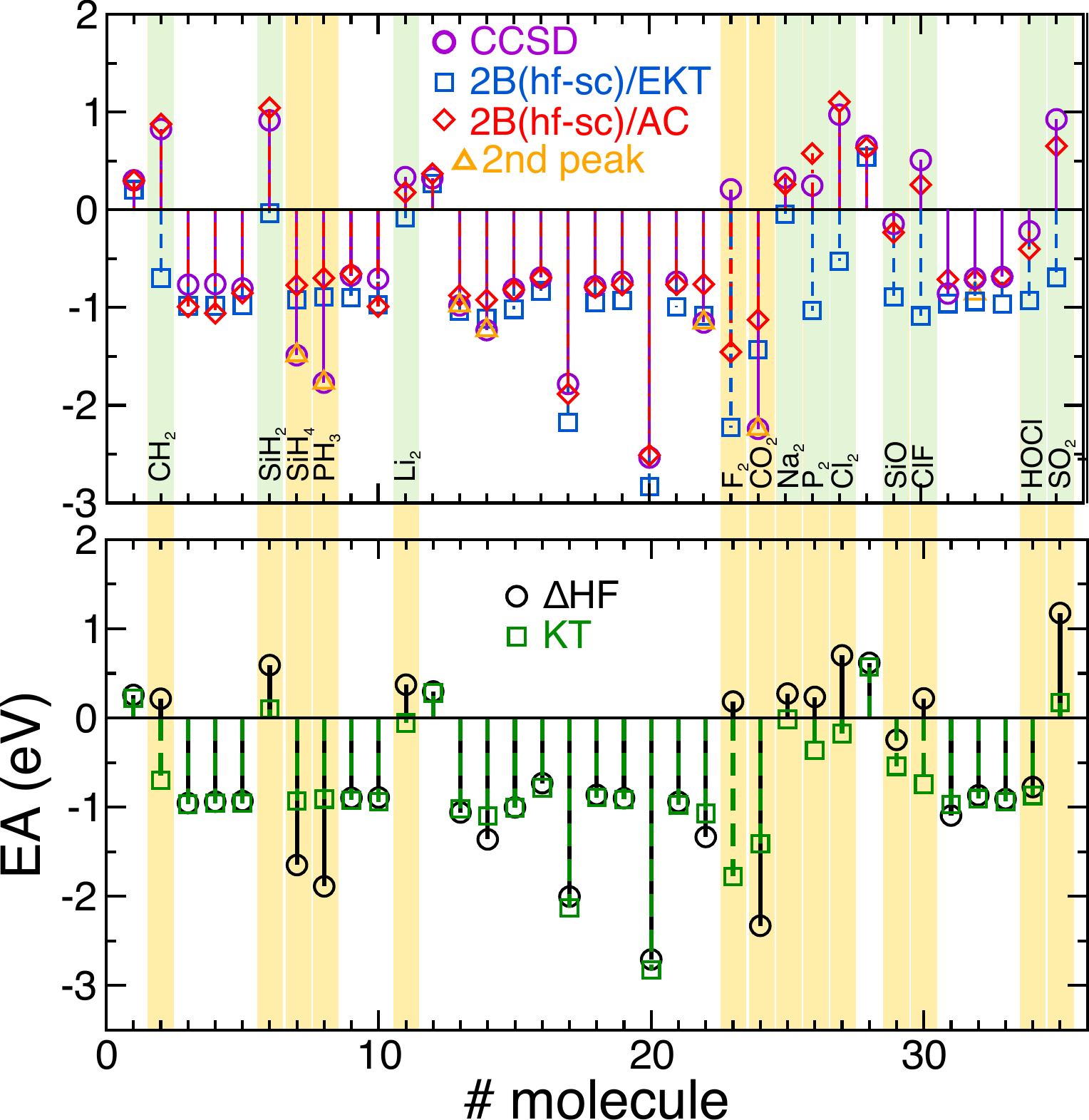}
  \caption{(Color online) Comparison of the electron affinities obtained from the
      HF self-consistent scheme using the extended Koopmans' theorem (EKT, blue squares)
      and the Pad\'{e} analytic continuation (AC, red diamonds) to the CCSD reference
    values. The orange triangles denote the second electron affinities. The yellow bars in
    the background indicate a significant discrepancy of $-\en_\mathrm{LUMO}$ (KT) to the
    EA computed by taking the difference of the total energy ($\Delta$HF). The cases with
    a good agreement between the EAs obtained from the Pad\'e AC and the CCSD
    reference are marked green background bars. \label{fig:ea}}
\end{figure}

In order to assess the performance of the 2BA, as compared to the reference methods, we
computed the IPs and compared them to the experimental values in
Fig.~\ref{fig:ip_vs_exp}. The EKT~\eqref{eq:EKT} was used to extract the IPs from the
MGF. Generally, the 2BA provides a quite accurate picture. Typical deviations from
experimental values, which occur within MP2 and CCSD, as well, are not cured by the
2BA. This can be related to the above mentioned multi-configuration problems. In
principle, these deficiencies can be rectified by starting from a multi-configurational HF
to construct the reference GF $\vec{g}(\tau)$.  As can already observed from the
distribution of the IPs in Fig.~\ref{fig:ip_vs_exp}(a)--(c), the non-sc scheme severely
overestimates the IPs. Comparing with the initial restricted Hartree-Fock (RHF)
values (which are mostly located under the diagonal), the non-sc treatment moves most of
the points up and thus "overshoots" the QP shifts. The HF-sc level, on the other hand,
yields much better results, which can be seen from the small distance of the points from
the diagonal. Visually, the predictions of the IPs by the HF-sc scheme is very similar to
the MP2 or CCSD reference. Switching to full self-consistency,
Fig.~\ref{fig:ip_vs_exp}(c), the values are slightly deteriorating with respect to the
HF-sc level. Such oscillatory behavior of the MBPT and the levels of self-consistency is
very typical. Similar behavior is also known for the $GW$ approximation, where partly
self-consistent schemes such as $G W_0$ or quasiparticle self-consistency are typically
superior to full-sc treatment.

For a quantitative analysis, we computed the mean absolute error (MAE) for each of the
methods:
\begin{align}
  \text{MAE}&=\frac{1}{N}\sum_{i=1}^{N}\left|E_i^\text{c}-E_i^\text{r} \right|.
\end{align}
Here the sum is performed over all systems, $E_i^\text{c}$ and $E_i^\text{r}$ refer to the
computed and the reference data, respectively. As inferred from
Fig.~\ref{fig:ip_vs_exp}(f), the 2BA on non-sc level is not even better than the RHF (IPs
from Koopmans' theorem), because of the overestimated QP shifts, while the accuracy of the
HF-sc scheme is comparable to the MP2 method. The quality of the full-sc treatment is on
the intermediate level between the RHF and the MP2. In principle, the 2BA is expected to
perform similarly to MP2, as both methods are of the second order in the Coulomb
interaction. Due to the oscillatory nature of MBPT theory~\cite{pavlyukh_pade_2017},
however, the partially self-consistent (HF-sc) level performs the best.

\begin{figure}[t]
  \centering
  \includegraphics[width=\columnwidth]{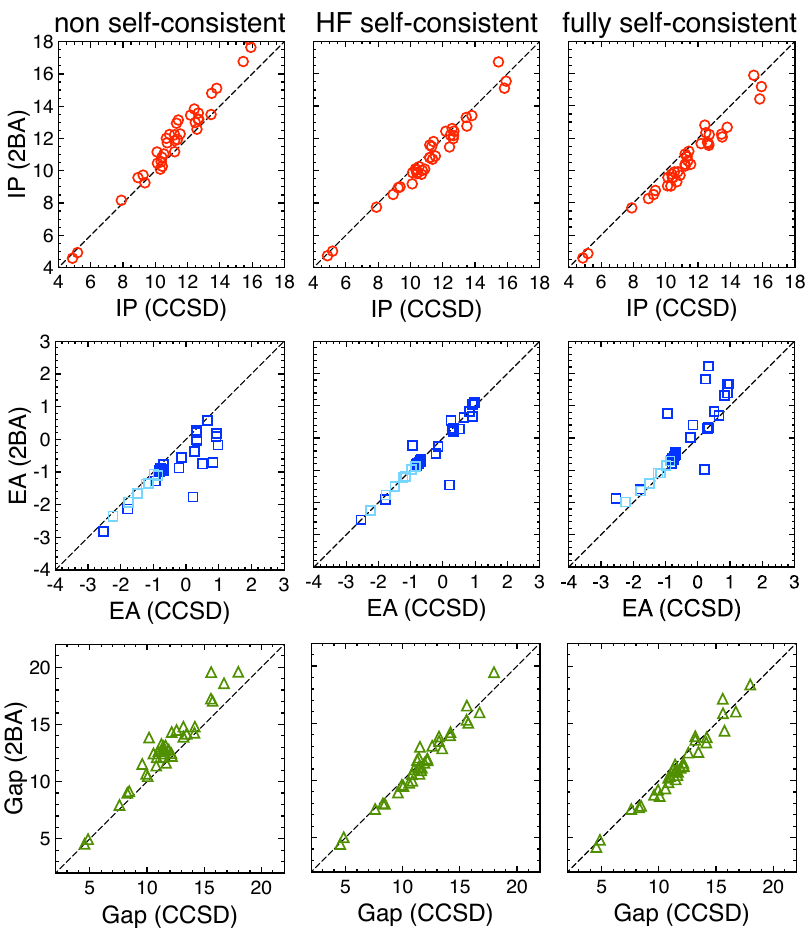}
  \caption{(Color online) Ionization potentials (top panels), electron affinities (middle
    panels) and band gaps (bottom panels) of the G2-1 molecules, obtained from the 2BA
    vs. the CCSD reference. Values are in eV. The Green's function calculation are
    performed using $E_\text{cut}=1.5$\,a.\,u. and the Pad\'{e}
    approximation. \label{fig:ip_ea_bg}}
\end{figure}

So far the IPs were computed using the EKT. As the next step we compared them to the IPs
extracted from the AC accomplished by the Pad\'{e} approximation
(Fig.~\ref{fig:ip_pade_vs_ekt}). Both methods yield almost identical values for
  non-sc case (left panel of Fig.~\ref{fig:ip_pade_vs_ekt}).  On the HF-sc and the full-sc
  levels, except for pathological cases such as the N$_2$, CS and H$_2$O$_2$ molecules,
  there are only some small deviations. N$_2$ and CS possess a triple bond leading to
stronger electron localization and thus electronic correlation, making such systems
difficult to treat. For instance, a diagrammatic expansion of the self-energy up to fourth
order is needed~\cite{day_electronic_2007} to correctly capture the orbital structure of
N$^+_2$. For a low-order diagrammatic approach, such strong many-body effects result in a
deviation from simple quasiparticle behavior which can not be captured by the EKT. Hence,
there are pronounced differences of the IP obtained by the AC and the
EKT. Similarly, the failure of Koopmans' theorem (KT) for CS due to pronounced correlation
effects is also known~\cite{ponzi_dynamical_2014}. Apart from such cases, the IPs obtained
by either method agree well.

The situation changes substantially for the electron affinities (Fig.~\ref{fig:ea}). For a
large class of molecules, the EKT provides a good estimate of the EA (we take CCSD as the
reference). However, for some molecules (CH$_2$, SiH$_2$, Li$_2$, F$_2$, CO$_2$, Na$_2$,
P$_2$, Cl$_2$, SiO, ClF, SO$_2$) the EA obtained by the EKT applied to the 2BA (HF-sc
level) is very different from the reference. These discrepancies are reminiscent of the
errors of the KT for the EAs within RHF. In fact, the EKT gives only small
QP shifts from the initial RHF energy levels entering the reference MGF
$\vec{g}(\tau)$. Hence, the EAs differ only little from $-\en_\mathrm{LUMO}$.  The above
molecules are typical cases where $-\en_\mathrm{LUMO}$ is a poor estimate for the EA (even
within the RHF).  Fig.~\ref{fig:ea} demonstrates that this behavior transfers to the EKT:
the molecules where the KT prediction differs substantially from the more accurate
estimation based on the total energy differences (the so-called $\Delta$HF method) are
identical to those where the EA obtained by the EKT is quite off the CCSD reference
value. However, employing the Pad\'{e} approximation yields a substantial improvement, as
the EAs obtained within the 2BA are much closer to the CCSD values. In particular, except
for the F$_2$ and P$_2$ dimers, the Pad\'{e} approximation always reproduces the correct
sign of the respective EAs. In cases where the modulus of the EAs is underestimated as
compared to CCSD, taking the second EAs (i.\,e. the second QP peak) leads to almost
perfect agreement. This is a clear indication of the multi-configurational instability of
the ground state of either the neutral or the negatively charged molecule. Such
deficiencies related to the HF starting point can, in principle, be overcome by the
full-sc treatment (as the dependence on the starting points disappears). However,
converging the Dyson equation towards the self-consistency can be hindered by the
multi-valuedness of the
solution~\cite{stan_unphysical_2015,tarantino_self-consistent_2017}.

Since several factors (besides the multi-configurational stability) contribute to the IPs
and EAs measured in experiments, accurate methods like the CCSD can, of course, not yield
perfect results. Most importantly, the restricted Gaussian basis set does describe excited
orbitals well. In order to compare the methods on equal grounds, we show the IPs, EAs, and
the resulting QP gap of the 2BA directly vs. the CCSD in Fig.~\ref{fig:ip_ea_bg}. As for
the IPs extracted by the EKT (Fig.~\ref{fig:ip_vs_exp}), one can infer that the HF-sc
scheme performs the best throughout; the agreement of the gaps between CCSD and the 2BA is
especially good. It does not rely on the errors cancellation for the electron affinities
(as illustrated in Fig.~\ref{fig:ea}, see important exceptions) and ionization potentials,
but is separately achieved for each quantity. Fig.~\ref{fig:ip_ea_bg} confirms that the
2BA on HF-sc level is almost comparable to the CCSD method.

We note that decreasing the basis cutoff to $E_\mathrm{cut} =0.5$~a.\,u. further
improves\,---\,as expected\,---\,the agreement of the IPs with the CCSD method. However,
the accuracy of the EAs is slightly deteriorated. The corresponding values are given in
Appendix~\ref{sec:app:G2-1}. The overall performance of the different levels of
self-consistency remains the same as for the larger cutoff.

\subsection{Ionization potentials and electron affinities of G2/97 molecules\label{subsec:g2-97}}

\begin{figure}[t]
  \centering
  \includegraphics[width=\columnwidth]{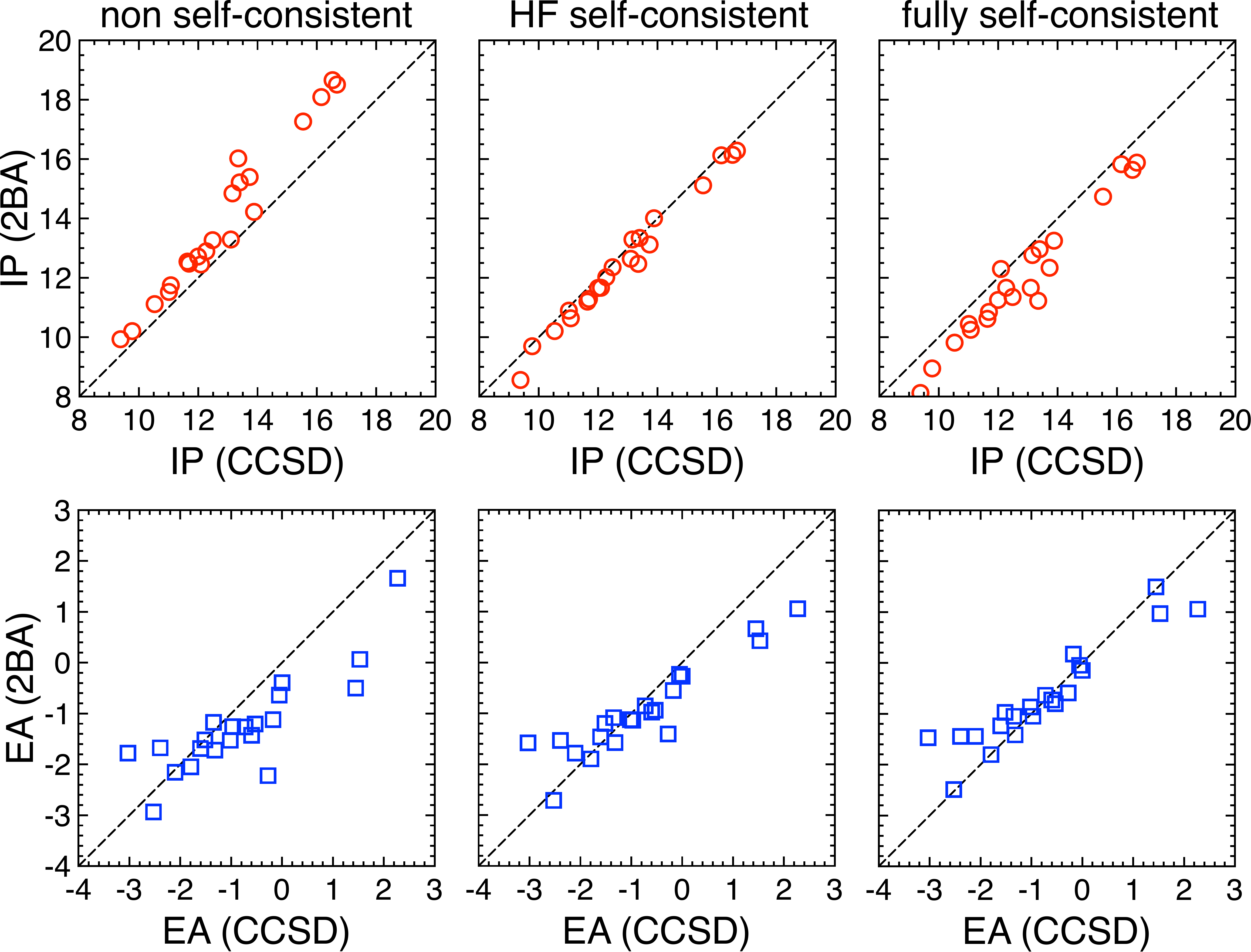}
  \caption{(Color online) Ionization potentials (top panels) and electron affinities
    (bottom panels) of the G2/97 molecules, obtained from the 2BA vs. the CCSD
    reference. Values are in eV. The Green's function calculation are performed
      using $E_\text{cut}=0.5$\,a.\,u. and the Pad\'{e}
      approximation. \label{fig:ip_ea_97}}
\end{figure}

Analogous to the G2-1 molecules, we have also performed calculations
for the non-hydrogenic closed-shell molecules from the G2/97 test
set. Since the molecules are composed by of heavier elements, the
number of valence orbitals and the HF basis increases considerably. In
Fig.~\ref{fig:ip_ea_97} the IPs and EAs computed using the Pad\'{e}
approximation and a basis determined by the cutoff
$E_\mathrm{cut}=0.5$~a.\,u. are presented agaist the results of the
CCSD method. Generally, a very good agreement is found.  Especially
the IP is well captured by the 2BA. Similarly to the results for the
G2-1 molecules, the partially self-consistent scheme performs best for
predicting the IP. Interestingly, the full-sc level improves the
accuracy of the EAs for this smaller basis size.  We have also tested
$E_\mathrm{cut}=1.5$~a.\,u. (see Appendix~\ref{sec:app:G2-97}), which
results in slightly better agreement of the EAs to the CCSD values at
HF-sc level at the cost of slightly decreasing the accuracy of the
IPs. For the full-sc scheme, on the other hand, both the IPs and the
EAs deviate more from the CCSD as for the smaller cutoff. In any case,
the HF-scheme performs best throughout and yields very good agreement
with the CCSD reference for both values of the cutoff.

\begin{figure}[b]
  \centering
  \includegraphics[width=\columnwidth]{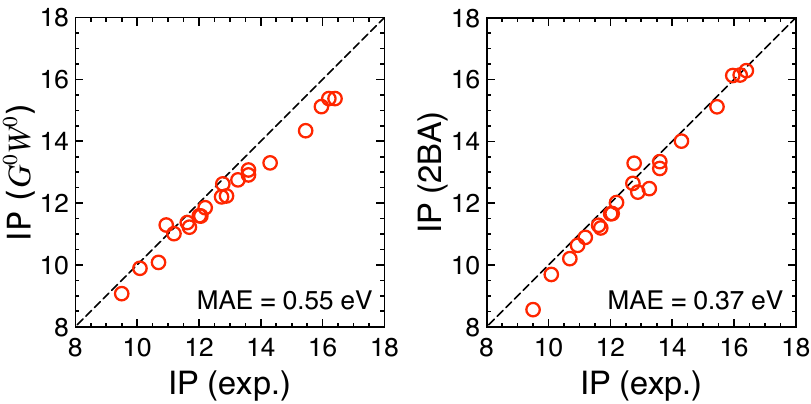}
  \caption{(Color online) Ionization potentials of the G2/97 molecules, comparing the
    $G^0W^0$ approximation from ref.~\onlinecite{pham_$gw$_2013} (left panel) and the
    HF-sc 2BA (right panel) to experimental values. Values are in eV. The Green's
      function calculation are performed using $E_\text{cut}=0.5$\,a.\,u. and the Pad\'{e}
      approximation. \label{fig:ip97_exp}}
\end{figure}

In Fig.~\ref{fig:ip97_exp} we compare the IPs within the 2BA treatment (HF-sc) and within
the $G^0W^0$ approximation (values taken from ref.~\onlinecite{pham_$gw$_2013}) to
experimental values. Both methods perform well; however, the MAE of 0.37~eV for the 2BA is
considerably smaller than the MAE of 0.55~eV obtained from the $G^0W^0$ approach. This
underlines that the 2BA is an excellent method for describing electronic properties of
molecules. 

\section{Conclusions\label{sec:summary}}
We performed benchmark calculations using the second Born approximation for the electron
self-energy of a number of molecular systems and found overall
very good performance of the Matsubara Green's function
approach in comparison with correlated quantum chemistry methods.  For all methods, the
error is substantially smaller if systems with multi-configurational ground state are
excluded. Same is true for $GW$ method~\cite{pavlyukh_configuration_2007}.

The partially self-consistent scheme has been demonstrated to perform the best throughout
with a predictive power on par with quantum chemistry methods. Nevertheless, the fully
self-consistent scheme performs very well for the majority of molecules, too. This is an
important requirement for performing time-dependent calculations, allowing to compute, for
instance, accurate optical absorption spectra.  An alternative possibility would be to
completely eliminate the Matsubara step and exploit the adiabatic switching scenario that
can be further facilitated by the use of generalized Kadanoff-Baym
Ansatz~\cite{galperin_linear_2008,pal_conserving_2009,
  ness_nonequilibrium_2011,latini_charge_2014}.

For extracting the quasiparticle properties we adopted two methods: the EKT and analytic
continuation. They yield almost identical results for electron removal energies, the EKT
was found to suffer from similar deficiencies as the KT within HF theory for around one
third of the investigated molecules. This drawback can be cured by analytic continuation
which yields excellent results for both ionization potentials and electron affinities.

As is mentioned in the introduction, there are several implementations of 2BA for systems
ranging from atoms~\cite{dahlen_self-consistent_2005}, to
Hubbard~\cite{puig_von_friesen_kadanoff-baym_2010,hopjan_merging_2016} and
Anderson~\cite{uimonen_comparative_2011} models, to ultracold
gases~\cite{schlunzen_dynamics_2016} and periodic
systems~\cite{rusakov_self-consistent_2016,welden_exploring_2016}, also addressing the
question of the accuracy of such calculations. Our study focuses on one missing aspect of
such studies, namely the performance of the method for molecular system, as a quick way to
initialize the time-dependent propagation of the Kadanoff-Baym equations. To this end, we
specifically focused on the inherent accuracy of 2BA.  For improving the general
predictive power, two key issues have to be adressed: an accurate solution of the Dyson
equation and systematic improvements of the basis. This first requirement is fulfilled by
our efficient solution scheme, based on a compact grid representation of the MGF. Further
improvements in this regard can be expected by working directly in the basis of orthogonal
polynomials or an optimized sparse
representation~\cite{shinaoka_compressing_2017,otsuki_sparse_2017}, which would allow
studying considerably larger systems or higher-order MBPT.
  Promising routes speeding up the calculation for larger systems while keeping a accurate
  single-particle basis is the already mentioned use of specialized basis to represent the
  self-energy~\cite{balzer_efficient_2010} or stochastic sampling
  methods~\cite{neuhauser_stochastic_2016}.

Finally we notice that similar to the quantum chemistry~\cite{kong_explicitly_2012} and
solid-state case~\cite{marsman_second-order_2009}, the explicitly correlated R12/F12
approaches are expected to recover even a larger portion of the correlation energy in the
Matsubara Green's function method. This idea opens new prospects for further
investigations.

\begin{acknowledgments}
  The calculations have been performed on the Beo04 cluster at the University of Fribourg
  and the Ianvs cluster at the Martin-Luther University Halle-Wittenberg. This work has
  been supported by the Swiss National Science Foundation through NCCR MARVEL and ERC
  Consolidator Grant No.~724103.  Y.P. acknowledges funding of his position by SFB/TRR
  173. We thank Hugo Strand, Philipp Werner and Michael van Setten for fruitful
  discussions.
\end{acknowledgments}

\appendix
\section{G2-1 molecules\label{sec:app:G2-1}}
In this section we present for completeness 2BA results for the molecules from the G2-1
set computed at the lower energy cutoff of $E_\text{cut}=0.5$\,a.u. using the Pad\'{e}
approximation, Tabs.~\ref{tab:ip:set:1:cutoff:05} and \ref{tab:ea:set:1:cutoff:05}. For
the CS molecule, the CCSD value for EA in aug-cc-pVDZ basis is not available; instead the
CCSD(T)/cc-pVDZ value from the NIST Standard Reference Database is
used~\cite{johnson_nist_nodate}. 
\renewcommand{\arraystretch}{1.1}
\begin{table}[b!]
\caption{\label{tab:ip:set:1:cutoff:05} Ionization potentials of the molecules from the
  G2-1 set. The energy cutoff of $E_\text{cut}=0.5$\,a.u. is used for the 2BA
  calculations.}
\begin{ruledtabular}
  \begin{tabular}{rrddddd}
    No.\vphantom{\Big[}&   System& \multicolumn{1}{c}{n-sc}& \multicolumn{1}{c}{HF-sc}&
    \multicolumn{1}{c}{full-sc}& \multicolumn{1}{c}{CCSD}& \multicolumn{1}{c}{Exp.}\\\hline
    1&   \ch{  LiH}&      8.17&      7.72&      7.65&      7.90&      7.90\\
    2&   \ch{  CH2}&     10.84&     10.06&      9.88&     10.41&     10.35\\
    3&   \ch{  NH3}&     11.73&     10.43&     10.28&     10.74&     10.07\\
    4&   \ch{ H2HO}&     13.84&     12.16&     11.96&     12.42&     12.62\\
    5&   \ch{   HF}&     17.64&     15.79&     15.56&     15.92&     16.03\\
    6&   \ch{ SiH2}&      9.25&      8.92&      8.76&      9.36&      8.92\\
    7&   \ch{ SiH4}&     13.16&     12.46&     12.35&     12.69&     11.00\\
    8&   \ch{  PH3}&     10.58&     10.00&      9.87&     10.36&      9.87\\
    9&   \ch{  H2S}&     10.47&      9.92&      9.72&     10.13&     10.46\\
   10&   \ch{  HCl}&     12.98&     12.41&     12.20&     12.48&     12.74\\
   11&   \ch{  Li2}&      4.92&      5.13&      2.62&      5.18&      5.11\\
   12&   \ch{  LiF}&     12.94&     10.90&     10.51&     11.31&     11.30\\
   13&   \ch{ C2H2}&     11.17&     10.95&     10.53&     11.20&     11.40\\
   14&   \ch{ C2H4}&     10.27&      9.70&      9.53&     10.44&     10.51\\
   15&   \ch{ C2H6}&     13.19&     12.25&     12.12&     12.65&     11.52\\
   16&   \ch{  HCN}&     13.48&     13.34&     12.87&     13.45&     13.70\\
   17&   \ch{   CO}&     15.12&     13.94&     13.48&     13.80&     14.01\\
   18&   \ch{ HCOH}&     12.01&     10.39&     10.05&     10.68&     10.88\\
   19&   \ch{CH3OH}&     12.25&     10.75&     10.53&     10.88&     10.84\\
   20&   \ch{   N2}&     16.73&     14.91&     14.30&     15.45&     15.58\\
   21&   \ch{ N2H4}&     11.15&      9.71&      9.51&     10.11&      8.10\\
   22&   \ch{ H2O2}&     13.16&     10.86&     10.39&     11.43&     10.58\\
   23&   \ch{   F2}&     18.15&     15.24&     14.60&     15.82&     15.70\\
   24&   \ch{  CO2}&     14.80&     13.09&     12.62&     13.49&     13.78\\
   25&   \ch{  Na2}&      4.56&      4.97&      2.69&      4.87&      4.89\\
   26&   \ch{   P2}&     10.12&     10.25&      9.47&     10.31&     10.53\\
   27&   \ch{  Cl2}&     12.28&     11.11&     10.78&     11.54&     11.48\\
   28&   \ch{ NaCl}&      9.57&      8.62&      8.46&      8.93&      9.20\\
   29&   \ch{  SiO}&     11.90&     10.71&     10.71&     11.37&     11.49\\
   30&   \ch{  CS} &     12.57&     10.98&     10.22&     12.58&     11.33\\
   31&   \ch{  ClF}&     13.58&     12.51&     12.19&     12.67&     12.74\\
   32&   \ch{Si2O6}&     11.01&     10.25&     10.04&     10.53&      9.74\\
   33&   \ch{CH3Cl}&     11.87&     11.00&     10.75&     11.20&     11.26\\
   34&   \ch{H3CSH}&      9.73&      9.00&      8.75&      9.24&      9.44\\
   35&   \ch{ HOCl}&     12.22&     10.86&     10.54&     11.20&     11.12\\
   36&   \ch{  SO2}&     13.45&     11.31&     11.03&     12.19&     12.35\\\hline
   \multicolumn{2}{c}{\textbf{MAE (eV)}}&
   0.99&      0.45&      0.80&      0.36
 \end{tabular}
\end{ruledtabular}
\end{table}
\begin{table}[b!]
\caption{\label{tab:ea:set:1:cutoff:05} Electron affinities of the molecules from the G2-1
  set. The 2BA calculations are performed at the energy cutoff of
  $E_\text{cut}=0.5$\,a.u.}
\begin{ruledtabular}
  \begin{tabular}{rrddddddd}
    No.\vphantom{\Big[}&   System&
    \multicolumn{1}{c}{n-sc}&
    \multicolumn{1}{c}{HF-sc}&
    \multicolumn{1}{c}{full-sc}&
    \multicolumn{1}{c}{CCSD}\\\hline
    1&   \ch{  LiH}&      0.23&      0.27&      0.30&      0.30\\
    2&   \ch{  CH2}&     -0.70&      0.10&     -0.82&      0.83\\
    3&   \ch{  NH3}&     -0.97&     -0.87&     -0.84&     -0.77\\
    4&   \ch{ H2HO}&     -0.96&     -0.88&     -0.85&     -0.76\\
    5&   \ch{   HF}&     -0.95&     -0.92&     -0.88&     -0.80\\
    6&   \ch{ SiH2}&      0.11&      0.76&      0.97&      0.91\\
    7&   \ch{ SiH4}&     -0.92&     -0.81&     -0.76&     -1.49\\
    8&   \ch{  PH3}&     -0.89&     -0.69&     -0.63&     -1.77\\
    9&   \ch{  H2S}&     -0.93&     -0.73&     -0.66&     -0.67\\
   10&   \ch{  HCl}&     -0.95&     -0.79&     -0.74&     -0.71\\
   11&   \ch{  Li2}&     -0.05&      0.22&     -0.43&      0.34\\
   12&   \ch{  LiF}&      0.27&      0.30&      0.32&      0.32\\
   13&   \ch{ C2H2}&     -1.02&     -0.88&     -0.80&     -0.98\\
   14&   \ch{ C2H4}&     -1.10&     -0.97&     -0.91&     -1.23\\
   15&   \ch{ C2H6}&     -1.00&     -0.88&     -0.83&     -0.81\\
   16&   \ch{  HCN}&     -0.79&     -0.69&     -0.61&     -0.69\\
   17&   \ch{   CO}&     -2.13&     -2.03&     -1.86&     -1.78\\
   18&   \ch{ HCOH}&     -0.90&     -0.84&     -0.77&     -0.78\\
   19&   \ch{CH3OH}&     -0.92&     -0.83&     -0.78&     -0.74\\
   20&   \ch{   N2}&     -2.88&     -2.73&     -2.65&     -2.53\\
   21&   \ch{ N2H4}&     -0.97&     -0.86&     -0.80&     -0.74\\
   22&   \ch{ H2O2}&     -1.07&     -0.96&     -0.88&     -1.15\\
   23&   \ch{   F2}&     -1.78&     -1.49&     -1.07&      0.21\\
   24&   \ch{  CO2}&     -1.41&     -1.23&     -1.04&     -2.24\\
   25&   \ch{  Na2}&      0.00&      0.29&      2.23&      0.33\\
   26&   \ch{   P2}&     -0.35&      0.17&      1.02&      0.25\\
   27&   \ch{  Cl2}&     -0.17&     -0.01&      0.21&      0.97\\
   28&   \ch{ NaCl}&      0.57&      0.59&      0.62&      0.66\\
   29&   \ch{  SiO}&     -0.56&     -0.39&     -0.01&     -0.14\\
   30&  \ch{CS}$^*$&     -1.27&     -0.77&     -0.18&     -0.93\\
   31&   \ch{  ClF}&     -0.75&     -0.53&     -0.32&      0.51\\
   32&   \ch{Si2O6}&     -0.96&     -0.74&     -0.63&     -0.86\\
   33&   \ch{CH3Cl}&     -0.90&     -0.77&     -0.70&     -0.70\\
   34&   \ch{H3CSH}&     -0.94&     -0.75&     -0.67&     -0.69\\
   35&   \ch{ HOCl}&     -0.87&     -0.70&     -0.63&     -0.22\\
   36&   \ch{  SO2}&      0.18&      0.32&      1.14&      0.93\\\hline
   \multicolumn{2}{c}{\textbf{MAE (eV)}}&
   0.44&      0.23&      0.49
 \end{tabular}
\end{ruledtabular}
\end{table}
\section{G2/97 molecules\label{sec:app:G2-97}}
In this section we present for completeness 2BA results for the molecules from the G2/97
set computed at two different energy cutoffs
$E_\text{cut}=0.5$\,a.u. (Tabs.~\ref{tab:ip:set:2:cutoff:05} and
\ref{tab:ea:set:2:cutoff:05}) and
$E_\text{cut}=1.5$\,a.u. (Tabs.~\ref{tab:ip:set:2:cutoff:15} and
\ref{tab:ea:set:2:cutoff:15}) using the Pad\'{e} approximation.  There are no reliable
experimental data for electron affinities of the molecules reported
Tabs.~\ref{tab:ea:set:2:cutoff:05} and \ref{tab:ea:set:2:cutoff:15}. Therefore, we use
CCSD results as reference for the computation of MAE.
\renewcommand{\arraystretch}{1.1}
\begin{table}[]
  \caption{\label{tab:ip:set:2:cutoff:15} Ionization potentials of the molecules from the
    G2/97 set. The energy cutoff of $E_\text{cut}=1.5$\,a.u. is used for the 2BA
    calculations.}
  \begin{ruledtabular}
    \begin{tabular}{rrdddddd}
      No.\vphantom{\Big[}&   System&
      \multicolumn{1}{c}{n-sc}&
      \multicolumn{1}{c}{HF-sc}&
      \multicolumn{1}{c}{full-sc}&
      \multicolumn{1}{c}{$G^0W^0$}&
      \multicolumn{1}{c}{CCSD}&
      \multicolumn{1}{c}{Exp.}\\\hline
    1&   \ch{  BF3}&     18.08&     15.35&     14.72&     15.12&     16.15&     15.96\\
    2&   \ch{ BCl3}&     12.48&     11.19&     10.43&     11.37&     11.69&     11.62\\
    3&   \ch{ AlF3}&     17.27&     14.54&     13.99&     14.34&     15.54&     15.45\\
    4&   \ch{AlCl3}&     12.73&     11.55&     10.96&     11.59&     12.00&     12.01\\
    5&   \ch{  CF4}&     18.65&     15.57&     14.79&     15.38&     16.53&     16.20\\
    6&   \ch{ CCl4}&     12.55&     11.03&     10.02&     11.22&     11.64&     11.69\\
    7&   \ch{  COS}&     11.54&     10.75&      9.90&     11.01&     11.01&     11.19\\
    8&   \ch{  CS2}&     10.21&      9.68&      8.55&      9.89&      9.78&     10.09\\
    9&   \ch{ CF2O}&     15.21&     12.52&     11.81&     12.91&     13.40&     13.60\\
   10&   \ch{ SiF4}&     18.52&     15.73&     15.09&     15.38&     16.68&     16.40\\
   11&   \ch{SiCl4}&     12.92&     11.57&      11.53&     11.58&     12.10&     12.06\\
   12&   \ch{  N2O}&     13.26&     11.93&     10.76&     12.23&     12.49&     12.89\\
   13&   \ch{ ClNO}&     11.76&     10.43&      8.77&     11.29&     11.08&     10.94\\
   14&   \ch{  NF3}&     15.39&     12.68&     11.71&     13.07&     13.73&     13.60\\
   15&   \ch{  PF3}&     12.89&     11.75&     11.20&     11.85&     12.27&     12.20\\
   16&   \ch{   O3}&     13.30&     10.85&      n.a.&     12.20&     13.10&     12.73\\
   17&   \ch{  F2O}&     16.02&     12.07&     10.83&     12.75&     13.35&     13.26\\
   18&   \ch{ ClF3}&     14.85&     11.98&     12.24&     12.62&     13.16&     12.77\\
   19&   \ch{ C2F4}&     11.12&      9.69&      8.85&     10.08&     10.53&     10.69\\
   20&   \ch{C2Cl4}&      9.92&      8.62&      7.42&      9.07&      9.38&      9.50\\
   21&   \ch{CF3CN}&     14.21&     13.55&     12.46&     13.30&     13.88&     14.30\\\hline
   \multicolumn{2}{c}{\textbf{MAE (eV)}}&
   1.14& 0.77 &1.50 & 0.55& 0.18
    \end{tabular}
  \end{ruledtabular}
\end{table}

\begin{table}
\caption{\label{tab:ea:set:2:cutoff:15} Electron affinities of the molecules from the
  G2/97 set. The energy cutoff of $E_\text{cut}=1.5$\,a.u. is used for the 2BA
  calculations.}
\begin{ruledtabular}
  \begin{tabular}{rrdddd}
    No.\vphantom{\Big[}&   System&
    \multicolumn{1}{c}{n-sc}&
    \multicolumn{1}{c}{HF-sc}&
    \multicolumn{1}{c}{full-sc}&
    \multicolumn{1}{c}{CCSD}\\\hline
    1   &\ch{  BF3}&     -1.18&     -1.00&     -0.85&     -1.35\\
    2   &\ch{ BCl3}&     -1.20&     -0.10&      0.88&     -0.53\\
    3   &\ch{ AlF3}&     -0.38&     -0.11&      0.12&      0.00\\
    4   &\ch{AlCl3}&     -0.65&     -0.02&      0.49&     -0.06\\
    5   &\ch{  CF4}&     -2.05&     -1.78&     -1.56&     -1.79\\
    6   &\ch{ CCl4}&     -1.42&     -0.47&      n.a.&     -0.60\\
    7   &\ch{  COS}&     -1.53&     -1.09&     -0.40&     -1.52\\
    8   &\ch{  CS2}&     -1.13&      0.19&      1.66&     -0.18\\
    9   &\ch{ CF2O}&     -1.67&     -1.32&     -1.00&     -2.39\\
   10   &\ch{ SiF4}&     -1.26&     -1.00&     -0.80&     -0.97\\
   11   &\ch{SiCl4}&     -1.25&     -0.69&      n.a.&     -0.72\\
   12   &\ch{  N2O}&     -2.14&     -1.49&     -0.64&     -2.10\\
   13   &\ch{ ClNO}&     -0.50&      1.62&     -2.54&      1.45\\
   14   &\ch{  NF3}&     -2.94&     -2.53&     -2.16&     -2.53\\
   15   &\ch{  PF3}&     -1.73&     -1.42&     -1.15&     -1.32\\
   16   &\ch{   O3}&      1.66&      2.22&      n.a.&      2.27\\
   17   &\ch{  F2O}&     -2.21&     -0.95&     -0.02&     -0.28\\
   18   &\ch{ ClF3}&      0.06&      1.26&      2.49&      1.53\\
   19   &\ch{ C2F4}&     -1.76&     -1.42&     -1.03&     -3.03\\
   20   &\ch{C2Cl4}&     -1.54&     -0.41&      n.a.&     -1.02\\
   21   &\ch{CF3CN}&     -1.70&     -1.30&     -0.85&     -1.60\\\hline
   \multicolumn{2}{c}{\textbf{MAE (eV)}}&
   0.67& 0.35 & 1.02
    \end{tabular}
  \end{ruledtabular}
\end{table}

\renewcommand{\arraystretch}{1.1}
\begin{table}[]
  \caption{\label{tab:ip:set:2:cutoff:05} Ionization potentials of the molecules from the
    G2/97 set. The energy cutoff of $E_\text{cut}=0.5$\,a.u. is used for the 2BA
    calculations.}
  \begin{ruledtabular}
    \begin{tabular}{rrdddddd}
      No.\vphantom{\Big[}&   System&
      \multicolumn{1}{c}{n-sc}&
      \multicolumn{1}{c}{HF-sc}&
      \multicolumn{1}{c}{full-sc}&
      \multicolumn{1}{c}{$G^0W^0$}&
      \multicolumn{1}{c}{CCSD}&
      \multicolumn{1}{c}{Exp.}\\\hline
    1&   \ch{  BF3}&     18.09&     16.13&     15.83&     15.12&     16.15&     15.96\\
    2&   \ch{ BCl3}&     12.48&     11.29&     10.85&     11.37&     11.69&     11.62\\
    3&   \ch{ AlF3}&     17.27&     15.12&     14.74&     14.34&     15.54&     15.45\\
    4&   \ch{AlCl3}&     12.72&     11.66&     11.25&     11.59&     12.00&     12.01\\
    5&   \ch{  CF4}&     18.66&     16.15&     15.64&     15.38&     16.53&     16.20\\
    6&   \ch{ CCl4}&     12.55&     11.20&     10.62&     11.22&     11.64&     11.69\\
    7&   \ch{  COS}&     11.52&     10.89&     10.44&     11.01&     11.01&     11.19\\
    8&   \ch{  CS2}&     10.20&      9.69&      8.95&      9.89&      9.78&     10.09\\
    9&   \ch{ CF2O}&     15.22&     13.34&     12.96&     12.91&     13.40&     13.60\\
   10&   \ch{ SiF4}&     18.51&     16.29&     15.88&     15.38&     16.68&     16.40\\
   11&   \ch{SiCl4}&     12.93&     11.66&     12.30&     11.58&     12.10&     12.06\\
   12&   \ch{  N2O}&     13.27&     12.36&     11.35&     12.23&     12.49&     12.89\\
   13&   \ch{ ClNO}&     11.75&     10.64&     10.25&     11.29&     11.08&     10.94\\
   14&   \ch{  NF3}&     15.40&     13.12&     12.34&     13.07&     13.73&     13.60\\
   15&   \ch{  PF3}&     12.89&     12.02&     11.66&     11.85&     12.27&     12.20\\
   16&   \ch{   O3}&     13.30&     12.64&     11.66&     12.20&     13.10&     12.73\\
   17&   \ch{  F2O}&     16.02&     12.47&     11.23&     12.75&     13.35&     13.26\\
   18&   \ch{ ClF3}&     14.85&     13.29&     12.77&     12.62&     13.16&     12.77\\
   19&   \ch{ C2F4}&     11.11&     10.21&      9.81&     10.08&     10.53&     10.69\\
   20&   \ch{C2Cl4}&      9.93&      8.56&      8.12&      9.07&      9.38&      9.50\\
   21&   \ch{CF3CN}&     14.23&     14.01&     13.25&     13.30&     13.88&     14.30\\\hline
   \multicolumn{2}{c}{\textbf{MAE (eV)}}&
   1.14&  0.37&  0.84 & 0.55 & 0.18
      \end{tabular}
  \end{ruledtabular}
\end{table}

\begin{table}[t]
\caption{\label{tab:ea:set:2:cutoff:05} Electron affinities of the molecules from the
  G2/97 set. The energy cutoff of $E_\text{cut}=0.5$\,a.u. is used for the 2BA
  calculations.}
\begin{ruledtabular}
  \begin{tabular}{rrdddd}
    No.\vphantom{\Big[}&   System&
    \multicolumn{1}{c}{n-sc}&
    \multicolumn{1}{c}{HF-sc}&
    \multicolumn{1}{c}{full-sc}&
    \multicolumn{1}{c}{CCSD}\\\hline
    1   &\ch{  BF3}&     -1.17&     -1.08&     -1.05&     -1.35\\
    2   &\ch{ BCl3}&     -1.21&     -0.94&     -0.80&     -0.53\\
    3   &\ch{ AlF3}&     -0.39&     -0.27&     -0.15&      0.00\\
    4   &\ch{AlCl3}&     -0.64&     -0.23&     -0.05&     -0.06\\
    5   &\ch{  CF4}&     -2.05&     -1.90&     -1.80&     -1.79\\
    6   &\ch{ CCl4}&     -1.43&     -0.98&     -0.73&     -0.60\\
    7   &\ch{  COS}&     -1.52&     -1.20&     -0.97&     -1.52\\
    8   &\ch{  CS2}&     -1.12&     -0.55&      0.17&     -0.18\\
    9   &\ch{ CF2O}&     -1.67&     -1.53&     -1.44&     -2.39\\
   10   &\ch{ SiF4}&     -1.26&     -1.14&     -1.05&     -0.97\\
   11   &\ch{SiCl4}&     -1.26&     -0.85&     -0.64&     -0.72\\
   12   &\ch{  N2O}&     -2.15&     -1.78&     -1.44&     -2.10\\
   13   &\ch{ ClNO}&     -0.50&      0.66&      1.49&      1.45\\
   14   &\ch{  NF3}&     -2.94&     -2.71&     -2.49&     -2.53\\
   15   &\ch{  PF3}&     -1.72&     -1.58&     -1.42&     -1.32\\
   16   &\ch{   O3}&      1.66&      1.06&      1.06&      2.27\\
   17   &\ch{  F2O}&     -2.22&     -1.41&     -0.59&     -0.28\\
   18   &\ch{ ClF3}&      0.07&      0.43&      0.97&      1.53\\
   19   &\ch{ C2F4}&     -1.78&     -1.58&     -1.47&     -3.03\\
   20   &\ch{C2Cl4}&     -1.53&     -1.13&     -0.87&     -1.02\\
   21   &\ch{CF3CN}&     -1.69&     -1.46&     -1.24&     -1.60\\\hline
   \multicolumn{2}{c}{\textbf{MAE (eV)}}&
   0.67 & 0.48 & 0.37
        \end{tabular}
  \end{ruledtabular}
\end{table}

\clearpage

%

\end{document}